\documentstyle[aps]{revtex}

\newcommand{\bra}[1]{\left\langle #1 \right|}
\newcommand{\ket}[1]{\left| #1 \right\rangle}

\begin{document}

\title{Asymmetric nuclear matter and neutron star properties}
\author{L.\ Engvik}
\address{Department of Physics,
University of Oslo, N-0316 Oslo, Norway}
\author{G.\ Bao}
\address{Department of Physics, AVH,
University of Trondheim, N-7055 Dragvoll, Norway}
\author{M.\ Hjorth-Jensen}
\address{ECT$^*$ European Centre for Theoretical Studies in
Nuclear Physics and Related Areas,
Trento, Italy}
\author{E.\ Osnes}
\address{Department of Physics,
University of Oslo, N-0316 Oslo, Norway}
\author{E. \O stgaard}
\address{Department of Physics, AVH,
University of Trondheim, N-7055 Dragvoll, Norway}

\maketitle

\begin{abstract}
In this work we calculate the total mass, radius, moment of inertia, and
surface gravitational redshift  for neutron
stars using various equations of state (EOS).
Modern meson-exchange potential models are used to
evaluate the $G$-matrix for asymmetric nuclear matter.
We calculate both a non-relativistic and a relativistic EOS.
Of importance here is the fact that relativistic Brueckner-Hartree-Fock
calculations for symmetric nuclear matter fit the empirical data, which
are not reproduced by non-relativistic calculations.
Relativistic  effects are known to be important at high densities,
giving an increased repulsion. This leads  to a stiffer EOS
compared to the EOS derived with a non-relativistic approach.
Both the non-relativistic and the relativistic EOS yield
moments of inertia and redshifts in agreement
with the accepted values. The relativistic EOS yields, however, too
large mass and radius.
The implications are discussed.
\end{abstract}

\section{Introduction}
The physics of compact objects like neutron stars offers
an intriguing interplay between nuclear processes  and
astrophysical observables (Pethick \& Ravenhall 1992;
Weber \& Glendenning 1993).
Neutron stars exhibit conditions far from those encountered on earth;
typically, expected densities $\rho$ of a neutron star interior are of the
order of $10^3$ or more times the density
$\rho_d\approx 4\cdot 10^{11}$ g/cm$^{3}$ at neutron ``drip''.
Thus, the
determination of an equation of state (EOS) for dense matter is
essential to calculations of neutron star properties; the EOS determines
properties  such as the mass range, the mass-radius
relationship, the crust thickness
and the cooling rate
(Weber \& Glendening 1993;
Pethick et al. 1995;
Lorenz et al. 1993).
The same EOS is also crucial
in calculating the energy released in a supernova explosion.
Clearly,
the relevant degrees of freedom will not be the same in the crust,
where the density is much smaller than the nuclear matter
saturation density, and in the center
of the star where the density is so high that models based
solely on interacting
nucleons are questionable.
Data from
neutron stars indicate that the EOS
should probably be moderately stiff in order to
support maximum neutron star masses in a range
from approximately
$1.4 M_{\odot}$ to $1.9 M_{\odot}$ (Thorsett et al. 93),

where $M_{\odot}$ is the solar mass. In addition,
simulations of supernova explosions seem to require an EOS which is
even softer. A combined analysis of the data coming from binary pulsar
systems and from neutron star formation scenarios is done
by Finn (1994), where it is shown that
neutron star masses should fall
predominantly in the range $1.3\le M/M_{\odot}\le 1.6 $.
In addition, a theoretical result
for the maximum mass of neutron stars will
have very important astrophysical
implications for the existence and number
of  black holes in the  universe;
examples are the famous galactic black
hole candidates Cyg X-1 (Gies \& Bolton 1986) and LMC X-3 (Cowley et al. 1983)
which are massive X-ray
binaries. Their mass functions (0.25 $M_{\odot}$
and 2.3 $M_{\odot}$) are, however, smaller
than for some low-mass X-ray binaries like A0620-00
(Mc Clintock \& Remillard 1986)
and V404 Cyg (Casares et al. 1992),
which make even better black hole candidates with mass functions in excess
of three solar masses.

Several theoretical approaches to calculations of the EOS
have been considered.
The hypothesis that strange quark matter may be the absolute
ground state of the strong interaction (Witten 1984),
has been used by Glendenning (1991) and Rosenhauer et al. (1992)
in the investigation of the possibility of interpreting pulsars as rotating
strange stars.
Other approaches introduce exotic states of nuclear matter,
such as kaon (Brown 1994; Thorsson et al. 1994; Brown et al. 1994;
Muto et al. 1993) or pion condensation
(Migdal et al. 1990; Takatsuka et al. 1993; Mittet \& \O stgaard 1988).

The scope of this work is to derive  the EOS
from the underlying many-body theory, derived
from a realistic  nucleon-nucleon (NN) interaction. From this EOS
we will study neutron-star observables such as mass-radius relationship,
surface gravitational redshift and moment of inertia.
By realistic we shall mean a nucleon-nucleon
interaction defined within the framework of meson-exchange theory,
 described conventionally in terms of one-boson-exchange (OBE) models
(Machleidt 1989;  Machleidt \& Li 1993). Explicitly, we will here build on the
Bonn meson-exchange potential models as they are defined by Macleidt (1989)
in table A.2.
Furthermore,
the physically motivated
coupling constants and energy cut-offs which determine the
OBE potentials are constrained through a fit to the
available  scattering data.
The subsequent step is to obtain an effective NN interaction in the
 nuclear
medium by solving the Bethe-Goldstone equation self-consistently.
Thus, the only parameters
which enter the theory are those which define the NN potential.
Until recently, most microscopic calculations of the EOS for
nuclear matter have been carried out within a
non-relativistic framework (Wiringa et al. 1988),
where the non-relativistic Schr\"{o}dinger equation is used to describe
the single-particle motion in the nuclear medium. Various degrees of
sophistication are accounted for in the literature
(Pethick \& Ravenhall 1992; Machleidt 1989)
ranging from first-order calculations in the reaction matrix $G$ to the
inclusion of two- and three-body higher-order
effects (Machleidt 1989; Wiringa et al. 1988; Dickhoff \& M\"{u}ther 1992;
Kuo \& Ma 1983; Kuo et al. 1987)

A common problem to non-relativistic nuclear matter calculations is,
 however, the
simultaneous reproduction of both the binding energy per nucleon
($BE/A=-16\pm 1$ MeV) and the saturation density with a Fermi momentum
$k_F=1.35\pm 0.05$ fm$^{-1}$.
Results obtained with a variety of methods and nucleon-nucleon (NN)
interactions are located along a band
denoted the ``Coester band'', which
does not satisfy the empirical data for nuclear matter.
Albeit these deficiencies, much progress has been achieved recently
in the description of the saturation properties of nuclear matter.
Of special relevance is the replacement of the non-relativistic
Schr\"{o}dinger equation with the Dirac equation to describe the
single-particle motion, referred to as the Dirac-Brueckner (DB)
 approach.
This is motivated by the success of the
phenomenological Dirac approach in nucleon-nucleus scattering
(Ray et al. 1991) and in the description of properties of finite nuclei
(Nikolaus et al. 1992), such as \ the spin-orbit splitting
in finite nuclei (Brockmann 1978). Moreover, rather promising results
within the framework of the DB approach have been obtained
(Brockmann \& Machleidt 1990; Li et al. 1992; M\"{u}ther et al. 1990),
 employing the
OBE models of the Bonn group. Actually, the empirical properties
of nuclear matter are quantitatively reproduced by
Brockmann \& Machleidt (1990).

Further, another scope
of this work is to derive the EOS for asymmetric nuclear
matter, using
both the non-relativistic
Brueckner-Hartree-Fock (BHF) approach and the relativistic
extension of this, the Dirac-Brueckner-Hartree-Fock (DBHF)
approach (Brockmann \& Machleidt 1990; Celenza \& Shakin 1986)
A preliminary discussion of
these results has been presented recently (Engvik et al. 1994).

We will treat the Pauli operator, which enters
our formalism (see below), correctly.
Asymmetric nuclear matter is important in e.g.\ studies of neutron
star cooling, as demonstrated recently by Lattimer et al. (1991).
They showed that ordinary nuclear matter
with a small asymmetry parameter can cool by the so-called
direct URCA process even more rapidly than matter in an exotic state.

However, as discussed above, an EOS with only nucleonic degrees
of freedom may not be too realistic in the crust region and at
high densities in the interior of the star. Thus, in our
study of neutron star observables we will have to link
our EOS with equations of state which take into account
degrees of freedom other than the nucleonic ones.

This work falls in four sections. After the
introductory remarks, we
briefly review the general picture in section 2.
In this section we also recast some of the astrophysical equations,
together with the equations of state derived within both the
non-relativistic and the Dirac-Brueckner-Hartree-Fock approaches.
The results are presented  in section 3, while discussions and
conclusions are given in section 4.

\section{General Theory}

In the interior of neutron stars, we find matter at densities
above the neutron "drip" $\rho_d\approx 4\cdot 10^{11}$ g/cm$^{3}$,
the density at which nuclei begin to dissolve and merge together,
and
the properties of  cold dense matter and the associated equation of
state are reasonably well understood at densities up to
$\rho_n\approx 3\cdot 10^{14}$g/cm$^{3}$.
In the high-density range  above $\rho_n$ the physical properties of matter
are still uncertain.

In the region between $\rho_d $ and $\rho_n$ matter
is composed mainly of nuclei,
electrons and free neutrons. The nuclei disappear at the upper end of this
density range because their binding energy decreases with increasing density.
The nuclei then become more neutron-rich and their stability decreases
until a critical value of the  neutron number
is reached, at which point the nuclei dissolve, essentially by merging
together. Since the nuclei present are very neutron-rich, the matter
inside nuclei is very similar to the free neutron gas outside.
However, the external neutron gas reduces the nuclear surface energy
appreciably, and it must vanish when the matter inside
nuclei becomes identical to that outside.

The free neutrons supply an increasingly larger fraction of the total pressure
as the density increases, but at neutron drip the pressure is almost entirely
due to electrons. Slightly above neutron drip the adiabatic index drops
sharply since the low-density neutron gas contributes appreciably to
the density but not much to the pressure, and it
does not rise again above 4/3 until
$\rho > 7\cdot 10^{12}$ g/cm$^3$. This means that no stable
stars can have central densities in this region.

Relatively "soft" equations of state have been proposed since the
average system energy is attractive at nuclear densities. However,
``stiff'' equations  of state may result from potentials for which the
average system interaction energy is dominated by the attractive
part of the potential at nuclear densities, but by the repulsive part at
higher densities. The stiffer equations of state give rise to important
changes in the structure and masses of heavy neutron stars. As the interaction
energy becomes repulsive above nuclear densities, the corresponding
pressure forces are better able to support stellar matter against
gravitational collapse. The result is that the maximum masses of stars
based on stiff equations of state are greater than those based on soft
 equations  of state. Also,  stellar models based on stiffer equations of state
have a lower central density, a larger radius and a thicker crust.
Such differences are important in determining mass limits for neutron stars,
their surface properties, moments of inertia, etc.

For low densities $\rho < \rho_n$,  where the nuclear force is expected to
be attractive, the pressure is softened somewhat by the inclusion of
interactions. For very high densities, however, the equation of
state is hardened due to the dominance of the repulsive core in
the nuclear potential.

At very high densities above $10^{15}$ g/cm$^3$,
 the composition is expected to include an appreciable number of hyperons
and the nucleon interactions  must be treated relativistically. Relativistic
many-body techniques for strongly interacting matter are, however, not fully
developed.  Presently developed nuclear equations  of state are also
subject to many uncertainties, such as the possibility
of neutron and proton superfluidity, of pion or kaon  condensation, of neutron
solidification, of phase transition to "quark matter'', and the
 consequences of the $\Delta$--nucleon resonance.

At densities significantly greater
than $\rho_n$, it is no longer
possible to describe  nuclear matter in terms of a non-relativistic
many-body Schr{\"o}dinger equation.
The "meson clouds''  surrounding the
nucleons begin to overlap and the picture of
localized individual particles interacting via two-body
forces becomes invalid. Even before this "break-down" different potentials
which  reproduce reliably low-energy phase shift data result in different
equations of state, since the potentials are sensitive to the repulsive
core region unaffected by the phase-shift data. If the fundamental
building blocks  of all strongly interacting particles are quarks,
any description of nuclear matter at very high densities should involve quarks.
When nuclei begin to ``touch'',  matter just above this
 density should undergo a phase transition at which quarks would
begin to "drip" out
of the nucleons and the result
would be quark matter, a degenerate Fermi liquid.

The main uncertainty  in neutron star models is the equation of state
of nuclear matter, particularly above typical  nuclear densities of $\rho\sim2.
8\cdot10^{14}$
g/cm$^3$. But our present understanding of the condensed matter
is already sufficient  to place quite strict limits  on masses
 and radii of stable neutron stars.

Neutron star models including realistic equations
 of state give the following general results:  Stars calculated with
 a stiff equation of state  have a lower central density, a larger radius,
and a much thicker crust than stars of the same mass computed from
a soft equation of state. Pion or kaon condensation and quark matter would
tend to contract neutron stars of a given mass  and decrease the
maximum mass.

Calculations give the following configurations in the interior:
The surface for $\rho< 10^6$ g/cm$^3$  is a region in which
temperatures and magnetic fields may affect the equation of state.
The outer crust for  $10^6$ g/cm$^3$ $< \rho < 4\cdot 10^{11}$g/cm$^3$
is a solid region where a Coulomb lattice of heavy nuclei coexist in
$\beta$-equilibrium  with a relativistic degenerate electron gas.
The    inner crust for
$4\cdot10^{11}$ g/cm$^3$ $< \rho < 2\cdot10^{14}$g/cm$^3$
consists of a lattice of neutron-rich nuclei together with a superfluid
 neutron gas and an electron gas. The neutron liquid for
 $2\cdot  10^{14}$ g/cm$^3$ $< \rho < 8\cdot10^{14}$g/cm$^3$ contains mainly
superfluid neutrons with a smaller concentration of superfluid
protons and normal electrons. The core region for $\rho> 8\cdot10^{14}$
g/cm$^3$ may not exist in some stars and will depend on the existence
of pion or kaon condensation, neutron solid, quark matter, etc.

The minimum mass of a stable neutron star  can be determined from  a minimum
in the  mass-radius relation M(R), or by
setting the mean value of the adiabatic index  $\Gamma$ equal to the
critical value
for radial stability against collapse. The resulting minimum mass is
$M\sim 0.1M_{\odot}$, where $M_\odot $ is the solar mass, with a
corresponding central density of $\rho\sim
10^{14}$ g/cm$^3$ and radius $R\sim 200 $ km.
The maximum mass equilibrium configuration is somewhat uncertain,
but all microscopic calculations give $M<2.7M_\odot$.
Note that the adiabatic index
criterion for the minimum mass is an approximate one, while the one involving
the M(R) condition is a precise one, if the mass limit is due to instability
of a radial mode.

Astronomical observations leading to global neutron star parameters
such as total mass, radius, or moment of inertia,  are important since they
are sensitive to microscopic model calculations.
The most reliable way of determining masses is
 via Kepler's third law in binary pulsars. Observations of such pulsars give
(approximately)
a common mass region consistent with all data of
$ 1.3 M_\odot < M< 1.9 M_\odot$. Present mass determinations for neutron
 stars are  all consistent with present stellar evolution theories.
There is, however, no firm evidence yet about the value for
the maximum mass of a neutron star.
A general limit for the maximum mass can
 be estimated by assuming the following: General
 relativity is the correct theory of gravitation. The equation of state must
 satisfy the "microscopic stability'' condition $dP/d\rho \geq 0$
and the causality condition
$dP/d\rho <c^2$, and should match
some known low-density  equation of state.
This gives an upper limit of $M\sim3-5
M_\odot$. Stiff equations of state in calculations predict
a maximum mass
in the range $M\sim1.5-2.7
M_\odot$. Rotation may
increase the maximum neutron star mass, but not appreciably, i.e., $< 20\%$.

We aim here at discriminating between
equations of state for asymmetric nuclear  matter
derived from non-relativistic and relativistic
approaches (to be discussed below).
As discussed above, relativistic effects become important at densities
higher than $\rho_n$, and it is therefore of interest to understand
whether the two approaches yield significantly different neutron star
properties. The derivation of the equations of state is discussed
in the first subsection, whereas the equations which define the
calculations of mass, radius, moment of inertia and gravitational
redshifts are discussed in the subsequent subsection.

\subsection{Derivation of the equation of state for asymmetric nuclear matter}

In this subsection we discuss both the non-relativistic and the relativistic
approach to the EOS in the framework of
the Brueckner-Hartree-Fock (BHF) theory.

\subsubsection{ Brueckner-Hartree-Fock approach for asymmetric nuclear matter}

Following the conventional many-body approach, we divide the full
hamiltonian $H=T+V$, with $T$ being the kinetic energy
and $V$ the bare NN potential,
into an unperturbed part $H_0 =T+U$ and an interacting part $H_I = V-U$,
such that
\[
   H=T+V=H_0 + H_I,
\]
where we have introduced an auxiliary single-particle (sp)  potential $U$. If
$U$ is chosen such that $H_I$ becomes small, then perturbative
many-body techniques can presumably be applied.
A serious obstacle to any perturbative treatment is the fact that the
bare NN potential $V$ is very large at short inter-nucleonic distances,
which renders a perturbative approach highly prohibitive. To overcome
this problem, we introduce the reaction matrix $G$ given
by the solution of
the Bethe-Goldstone equation (in operator form)
\begin{equation}
   G(E)=V+VQ\frac{1}{E - QH_0Q}QG,
   \label{eq:bg}
\end{equation}
where $E$ is the energy of the interacting  nucleons and
 $Q$ is the the Pauli operator which prevents scattering into occupied states.
The Pauli operator is given by
\begin{equation}
   Q(k_m\tau_m , k_n \tau_n) =
    \left\{\begin{array}{cc}1,&k_m >k_F^{\tau_m} ,k_n>k_F^{\tau_n}  ,\\
    0,&otherwise,\end{array}\right.
    \label{eq:pauli}
\end{equation}
in the laboratory system, where $k_F^{\tau_i}$ defines the Fermi
momenta of the
proton ($\tau_i=1/2$) and neutron ($\tau_i=-1/2$).
For notational economy, we set $|{\bf k}_m|=k_m$.

The above expression for the Pauli operator is in the
laboratory frame. In the calculations of the $G$-matrix,
we will employ a Pauli operator in the center of mass and relative
coordinate system. Further, this Pauli operator
will be given by the so-called angle-average approximation,
for details see Bao et al. (1994).  Eq. (\ref{eq:bg}) reads
then (in a partial wave representation)
\begin{equation}
   G_{ll'}^{\alpha T_z}(kk'KE)=V_{ll'}^{\alpha T_z}(kk')
   +\sum_{l''}\int \frac{d^3 q}{(2\pi )^3}V_{ll''}^{\alpha T_z}(kq)
   \frac{Q^{T_z}(q,K)}{E-H_0}
   G_{l''l'}^{\alpha T_z}(qk'KE),
   \label{eq:gnonrel}
\end{equation}
with $ll'$ and $kk'$ the orbital angular momentum and the linear
momentum of the
relative motion, respectively. The angle-average Pauli operator
is given by $Q^{T_z}$, where $T_z$ is the total isospin projection.
Further $K$ is the momentum of the center-of-mass
motion. Since we are going to use an angular
average for the Pauli operator, the $G$-matrix is diagonal in total
angular momentum $J$.
Further, the $G$-matrix is diagonal in the center-of-mass orbital momentum
$L$ and the total spin $S$.
These quantities, i.e. $J$,  $L$ and $S$, are all represented
by the variable $\alpha$. Three different G-matrices  have to be evaluated,
depending of the individual isospins ($\tau_1\tau_2$)  of the interacting
nucleons
($\frac{1}{2}\frac{1}{2}$, $-\frac{1}{2}-\frac{1}{2}$ and
$-\frac{1}{2}\frac{1}{2})$.
These quantities are represented by the total isospin projection $T_z$
in eq. (\ref{eq:gnonrel}).
The different G-matrices  originate from the discrimination
between protons and neutrons in eq. (\ref{eq:pauli}).
The term $H_0$ in the denominator of eq. (\ref{eq:gnonrel}) is
the unperturbed energy of the intermediate states and depends
on $k$, $K$ and the individual isospin of the interacting particles; further
discussion is given below.

We use a continuous single particle (sp) spectrum advocated by
Mahaux et al. (1989). It is  defined by the
self-consistent solution of the following equations\footnote{
In this work we will throughout set $G=c=\hbar=1$, where
$G$ is the gravitational constant.}:
\begin{equation}
   \varepsilon_i    = t_i + u_i=\frac{k_i^2}{2m}
   +u_i,
   \label{eq:spnrel}
\end{equation}
where $m$ is the bare nucleon mass,
and
\begin{equation}
   u_i = \sum_{h\leq k_F}
   \left\langle i h \right|
  G(E = \varepsilon_{i} + \varepsilon_h )
   \left|  ih \right\rangle _{\mathrm{AS}}.
   \label{eq:selfcon}
\end{equation}
In eqs.\ (\ref{eq:spnrel})-(\ref{eq:selfcon}), the subscripts $i$ and $h$
represent the quantum numbers of the single-particle states, such as
isospin projections $\tau_i$ and $\tau_h$, momenta $k_i$ and $k_h$ etc.
The sp kinetic energy is given by $t_{i}$ and similarly the sp potential by
$u_{i}$. For further details see Bao et al. 1994.

Finally, the non-relativistic
energy per particle ${\cal E}/A$ is formally given as
\begin{eqnarray}
   {\cal E}/A &= &
    \frac{1}{A} \sum_{h\leq k_F}
    \frac{k_h^2}{2m}+ \\
 & &   \frac{1}{2A}
     \sum_{h \leq k_F,h'\leq k_F}
   \bra{hh'}G(E=\varepsilon_h +\varepsilon_{h'})
   \ket{hh'}_{\mathrm{AS}}  \nonumber
   \label{eq:enrel}
\end{eqnarray}
In this equation  we have suppressed the isospin indeces for the Fermi momenta.

\subsubsection{Relativistic effects}

The properties of neutron stars depend on the equation of state
at densities up to an order of magnitude higher than those observed
in ordinary nuclei. At such densities, relativistic effects
certainly prevail. Among relativistic approaches to the nuclear
many-body problem, the so-called Dirac-Hartree and Dirac-Hartree-Fock
approaches have received much interest
(Serot \& Walecka 1986; Serot 1992; Horowitz \& Serot 1987).
One of the early successes of these
approaches was the quantitative reproduction of spin observables,
which are only poorly described by the non-relativistic theory.
Important to these methods
was the introduction of a strongly attractive scalar component
and a repulsive vector component
in the nucleon self-energy (Brockmann 1978; Serot \& Walecka 1986).
Inspired
by the successes of the Dirac-Hartree-Fock method, a relativistic
extension of Brueckner theory was proposed by Celenza \& Shakin (1986),
known as the Dirac-Brueckner theory.
One of the appealing features of the Dirac-Brueckner approach
is the self-consistent determination of the relativistic sp
energies and wave functions.
The Dirac-Brueckner approach differs from the Dirac-Hartree-Fock one
in the sense that in the former one starts from the free NN potential
which is only constrained by a fit to the NN data, whereas the
Dirac-Hartree-Fock method pursues a line where the parameters of the
theory are determined so as to reproduce the bulk properties
of nuclear matter. It ought, however, to be stressed that the Dirac-Brueckner
approach,
which starts from NN potentials based on meson exchange,
is a non-renormalizable theory where short-range part
of the potential depends on additional parameters like
vertex cut-offs, clearly minimizing the
sensitivity of calculated results to short-distance inputs
(Brockmann \& Machleidt 1990; Celenza \& Shakin 1986; ter Haar \& Malfliet
1987).
The description presented here for the Dirac-Brueckner approach follows
closely that of Brockmann \& Machleidt (1990). We will thus
use the meson-exchange models of the Bonn group, defined
in table A.2, Machleidt (1989). There the three-dimensional
reduction of the Bethe-Salpeter equation as given by the Thompson equation
is used to solve the equation for the scattering
matrix (Thompson 1970). Hence, including the necessary medium effects like the
Pauli operators discussed in the previous subsection and the starting
energy, we will rewrite eq. (\ref{eq:gnonrel}) departing from
the Thompson equation. Then, in a self-consistent way, we
determine the above-mentioned scalar and vector components  which
define the nucleon self-energy.

In order to introduce the relativistic nomenclature, we consider first the
Dirac equation for a free nucleon, i.e.,
\[
  (i\not \partial -m )\psi (x)=0,
\]
where $m$ is the free nucleon mass
and $\psi (x)$ is  the nucleon field operator
($x$ is a four-point)
which is conventionally expanded in terms of plane wave states
and the Dirac spinors $u(p,s)$, and $ v(p,s)$, where
 $p=(p^0 ,{\bf p})$ is
a four momentum\footnote{Further notation is  as
given  in Itzykson  Zuber (1980).}
 and  $s$ is the spin projection.

The positive energy Dirac spinors are
(with $\overline{u}u=1$)
\begin{equation}
   u(p,s)=\sqrt{\frac{E(p)+m}{2m}}
	  \left(\begin{array}{c} \chi_s\\ \\
	  \frac{\mbox{\boldmath $\sigma$}\cdot{\bf p}}{E(p)+m}\chi_s
	  \end{array}\right),
   \label{eq:freespinor}
\end{equation}
where $\chi_s$ is the Pauli spinor and $E(p) =\sqrt{m^2 +|{\bf p}|^2}$.
To account for medium modifications to the free Dirac equation,
we introduce the notion of the self-energy $\Sigma (p)$.
As we assume parity to be a good quantum number, the self-energy of a
nucleon can be formally written as
\[
       \Sigma(p) =
       \Sigma_S(p) -\gamma_0 \Sigma^0(p)
       +\mbox{\boldmath $\gamma$}{\bf p}\Sigma^V(p).
\]
The momentum dependence of $\Sigma^0$
and $\Sigma_S$ is rather weak (Serot \& Walecka 1986).
Moreover, $\Sigma^V << 1$, such
that the features of the Dirac-Brueckner-Hartree-Fock
procedure can be discussed within the framework of the phenomenological
Dirac-Hartree ansatz, i.e. we  approximate
\[
\Sigma \approx \Sigma_S -\gamma_0 \Sigma^0 = U_S + U_V,
\]
where $U_S$ is an attractive
scalar field and $U_V$ is the time-like component
of a repulsive vector field.
The finite self-energy modifies the
free Dirac spinors of eq. (\ref{eq:freespinor}) as
\[
   \tilde{u}(p,s)=\sqrt{\frac{\tilde{E}(p)+\tilde{m}}{2\tilde{m}}}
	  \left(\begin{array}{c} \chi_s\\ \\
	  \frac{\mbox{\boldmath $\sigma$}\cdot{\bf p}}
          {\tilde{E}(p)+\tilde{m}}\chi_s
	  \end{array}\right),
\]
where we let the terms with tilde
represent the medium modified quantities.
Here we have defined
\[
   \tilde{m}=m+{U_S},
\]
and
\begin{equation}
      \tilde{E}_{i}=
      \tilde{E}(p_{i})=\sqrt{\tilde{m}^2_i+{\bf p}_{i}^2}.
\label{eq:eirel}
\end{equation}
As in the previous subsection, the subscripts $i$, and $h$ below,
represent the quantum numbers of the single-particle states, such as
isospin projections $\tau_i$ and $\tau_h$, momenta $k_i$ and $k_h$, etc.

The sp energy is
\begin{equation}
   \tilde{\varepsilon}_{i} =\tilde{E}_{i} +U_V^i,
   \label{eq:sprelen}
\end{equation}
and the sp potential is given by the G-matrix as
\begin{equation}
   u_{i} =\sum_{h\leq k_F}
         \frac{ \tilde{m}_i\tilde{m}_h }{ \tilde{E}_h \tilde{E}_{i} }
	\bra{i h}\tilde{G}(\tilde{E}=\tilde{E}_i +\tilde{E}_h)\ket{i h}_{\mathrm{AS}},
	\label{eq:urel}
\end{equation}
or, if we wish to express it in terms of the constants $U_S$ and
$U_V$,
we have
\begin{equation}
   u_{i} = \frac{\tilde{m}_i}{\tilde{E}_{i}}{U_S}^{i} +U_V^{i}.
   \label{eq:sppotrel}
\end{equation}
In eq. (\ref{eq:urel}), we have introduced the relativistic
$\tilde{G}$-matrix. If the two interacting particles ,
with isospins $\tau_1$ and $\tau_2$, yield a total isospin projection
$T_z$, the relativistic $\tilde{G}$-matrix
in a partial wave representation is given by
\begin{equation}
   \tilde{G}_{ll'}^{\alpha T_z}(kk'K\tilde{E})=
   \tilde{V}_{ll'}^{\alpha T_z}(kk')
   +\sum_{l''}\int \frac{d^3 q}{(2\pi )^3}
   \tilde{V}_{ll''}^{\alpha T_z}(kq)
   \frac{\tilde{m}_1 \tilde{m}_2} { \tilde{E}_1^{q} \tilde{E}_2^{q} }
   \frac{Q^{T_z}(q,K)} {( \tilde{E}-\tilde{E}_1^{q}-\tilde{E}_2^{q})}
   \tilde{G}_{l''l'}^{\alpha T_z}(qk'K\tilde{E}),
   \label{eq:grel}
\end{equation}
where the relativistic starting energy
is defined according to eq. (\ref{eq:eirel}) as
\[\tilde{E}=\tilde{E}(\sqrt{k^2+K^2/4},\tau_1)+
\tilde{E}(\sqrt{k^2+K^2/4},\tau_2).\]
and
\[\tilde{E}_{1 (2)}^{q}=\tilde{E}(\sqrt{q^2+K^2/4},\tau_{1(2)})\]

Equations (\ref{eq:sprelen})-(\ref{eq:grel}) are solved self-consistently,
starting
with adequate values for the scalar and vector components
$U_S$ and $U_V$. This iterative scheme is continued until these
parameters show little variation. The calculations are carried out in the
neutron matter rest frame, avoiding thereby a cumbersome
transformation between the two-nucleon center-of-mass system
and the neutron matter rest frame.

Finally, the relativistic version of eq. (\ref{eq:enrel}) reads
\begin{eqnarray}
   {\cal E}/A &=&
   \frac{1}{A}\sum_{h\leq k_F}
   \frac{ \tilde{m_h}m+ k_h^2} {\tilde{E}_h}+ \\
& &   \frac{1}{2}\sum_{h \leq k_F,h'\leq k_F}
   \frac{\tilde{m}_h\tilde{m}_{h'}}{\tilde{E}_h\tilde{E}_{h'}}
   \bra{hh'}\tilde{G}
   (\tilde{E}=
    \tilde{E}_h +\tilde{E}_{h'})
    \ket{hh'}_{\mathrm{AS}}
    -m.  \nonumber
   \label{eq:erel}
\end{eqnarray}

\subsection{Neutron star equations}
We end this section by presenting the formalism needed in order
to calculate the mass, radius, moment of inertia and gravitational
redshift.
We will assume that the neutron stars we study exhibit an isotropic
mass distribution. Hence, from the general theory of relativity,
the structure of a neutron star is determined through the
Tolman-Oppenheimer-Volkov eqs., i.e.,
\begin{equation}
   \frac{dP}{dr}=
		 - \frac{\left\{\rho (r)+P(r) \right\}
		  \left\{M(r)+4\pi r^3 P(r)\right\}}{r^2- 2rM(r)},
   \label{eq:tov}
\end{equation}
and
\begin{equation}
   \frac{dM}{dr}=
		 4\pi r^2 \rho (r),
   \label{eq:derM}
\end{equation}
where $P(r)$ is the pressure, $M(r)$ is
the gravitational mass inside a radius $r$, and $\rho (r)$ is the
mass-energy density. The latter equation is conventionally written
as an integral equation
\begin{equation}
   M(r)=
	4\pi \int_{0}^{r} \rho (r')r'^{2} dr' .
   \label{eq:M}
\end{equation}
In addition, the  main ingredient in a calculation of astrophysical
observables is the equation of state (EOS)
\begin{equation}
   P(n)=
     n^2 \left(\frac{\partial \epsilon}{\partial n}\right),
  \label{eq:P}
\end{equation}
where $\epsilon ={\cal E}/A$ is
the energy per particle and $n$ is the particle number
density.
Eqs.\ (\ref{eq:tov}), (\ref{eq:M}) and (\ref{eq:P}) are the basic
ingredients in our calculations of neutron star properties.

The moment of inertia $I$ for a slowly rotating symmetric neutron star
is related to the angular momentum $J$
and the angular velocity $\Omega$ in an
inertial system at infinity through
\begin{equation}
  I=
    \left(\frac{\partial J}{\partial \Omega}\right)_{\Omega =0}=
    \frac{J}{\Omega}.
    \label{eq:I}
\end{equation}
The angular momentum $J$ is given by
\begin{equation}
   J=\frac{1}{6}R^{4}\left(\frac{d\tilde{\omega}}{dr}\right)_{r=R},
\end{equation}
where
$\tilde{\omega}$ is the angular velocity relative to particles with zero
angular momentum.
Further, the angular velocity $\Omega$ is
\begin{equation}
   \Omega = \tilde{\omega}(R)+\frac{u(R)}{3R^3},
\end{equation}
with
\begin{equation}
   u=r^4 \frac{d\tilde{\omega}}{dr}.
\end{equation}

Finally, the gravitational redshift $Z_s$ is given by
\begin{equation}
Z_s=\left(1-\frac{2M(R)}{R}\right)^{-1/2}-1.
\end{equation}

To calculate the total mass, radius, moment of inertia and gravitational
redshift, we employ the EOS defined in eq. (\ref{eq:P}) with the
boundary conditions
\[
  P_c =P(n_c ), \hspace{1cm} M(0)=0,
\]
where we let the subscript $c$ refer to the center of the star, and
$n_c$ is the central density which
is our input parameter in the  calculations
of neutron star properties.

\section{Results}

\subsection{The equation of state}

As mentioned in the introduction,
the replacement of the non-relativistic Schr\"{o}dinger equation by the
Dirac equation offers a quantitative reproduction of the saturation
properties of nuclear matter. Central to these results is the
use of modern meson-exchange potentials with a weak tensor force, where the
strength of the tensor force is reflected in the $D$-state probability
of the deuteron. The main differences in the strength of the tensor force
in nuclear matter arises in the $T=0$ $^3S_1$-$^3D_1$ channel,
though other partial waves also give rise to tensor force contributions.
To derive the equation of state, we  start from the Bonn NN potential
models as they are defined by the parameters of table A.2,
Machleidt (1989)
These potentials are recognized by the labels A, B and
C, with the former carrying the weakest tensor force. Since all three
potentials
have to reproduce the same set of scattering data, a potential containing a
weak
tensor component needs a strong central component.
Due to the fact that the  tensor force is more quenched than the central force
in a medium,
the three potentials have different off-shell properties.
The stronger off-shell  interaction
of version A give rise to more binding energy in symmetric nuclear matter.

In neutron matter ($T=1$), however,
the important $^3S_1$-$^3D_1$ channel
does not contribute to the energy per particle, and the difference between
the various potentials is found  to be small.
This is indeed the case,
as reported in neutron matter calculations
(Li et al. 1992; Bao et al. 1994).
As the proton fraction increases, the $T=0$ contribution becomes more important
and the difference between the potentials becomes significant.
In our presentation we have used version A, since this potential reproduces the
empirical properties
of nuclear matter in the Dirac-Brueckner approach.
The energy shift going from symmetric nuclear matter
to pure neutron matter is largest for potential A, since A has the strongest
off-shell $T=0$ interaction.

In fig. \ref{fig:energy} we
present the results for both relativistic and non-relativistic
calculations. The relativistic effects become
significant at normal nuclear matter
densities.
The relativistic calculations give an increased
repulsion at higher densities
and correspondingly stiffer EOS than the non-relativistic
approaches.

Independent of the proton fraction the relativistic effects become
 significant at about 0.15 fm$^{-3}$.
Moreover, the energy shift  due to the relativistic effects
is almost the same for all different proton fractions.
The energy shift is only
a few MeV larger for neutron matter than for symmetric nuclear matter.

The differences between the relativistic and non-relativistic results
can be understood from the following two arguments,
see e.g. Brown et al. (1987).
Firstly, relativistic effects introduce a strongly
density dependent repulsive term in the energy per particle,
of the order of $(n/n_0)^{8/3}$, where $n_0$ is the
nuclear matter saturation density in fm$^{-3}$.
This contribution\footnote{The relativistic effects can also be understood as a
special class of many-body forces.}
is important in order to saturate nuclear matter, and is interpreted by
Brown et al. (1987)
as a density dependent correction
to the mass of the scalar boson $\sigma$, which
is responsible for the scalar term $U_S$ in the relativistic nucleon mass.
In the
vacuum, the $\sigma$-meson has self-energy contributions due to its
coupling to virtual nucleon-antinucleon pairs. In nuclear matter,
scattering into states with $k<k_F$ are Pauli blocked, giving in turn
a repulsive contribution to the energy per particle.
Secondly, the nucleon-nucleon spin-orbit interaction is enhanced
(the spin-orbit force is repulsive), since the
relativistic effective mass is changed due to the scalar fields which
couple to negative energy states. For further details on the relativistic
effects, see the discussion in chapter 10 of Machleidt (1989).

For relatively small proton fractions,
the energy per
particle exhibits much the same curvature as the curve for pure neutron
matter at high densities, although the energy per particle is less repulsive
at high densities.
At lower densities, the situation is rather
different.
This is due to the contributions from various isospin $T=0$ partial waves,
especially the contribution from the $^3S_1$-$^3D_1$ channel, where the
tensor force component of the nucleon-nucleon potential provides
additional binding.
{}From fig.\ \ref{fig:energy} we note that
with a proton fraction of $15\%$, the energy per particle
starts to become attractive at low densities (in the region $0.07$ fm$^{-3}$
to $0.3$ fm$^{-3}$). For larger proton fractions, the attraction is  increased.

Using the non-relativistic and the relativistic
equations of state, we wish to study how sensitive various neutron star
properties
are with respect to different proton fractions.
Note also that within the
Dirac-Brueckner approach, the Bonn A potential reproduces the empirical nuclear
matter
binding energy and saturation density (Brockmann \& Machleidt 1990).
This gives a more
consistent approach to asymmetric nuclear matter. The reader should, however,
keep in mind that there are several mechanisms (to be discussed in
section 4) which may reduce the stiffness of the above equations of state.

\subsection{Mass, radius, moment of  inertia and  surface gravitational
redshift}

To calculate mass, radius, moment of  inertia and  surface gravitational
redshift we need the EOS for all relevant densities. The equations
of state derived in the previous subsection have a limited range,
$0.1$ fm$^{-3} \leq n \leq 0.8$  fm$^{-3}$. We
must therefore include equations of state for other densities as well.
These equations of state are discussed below.

For the lowest densities, we use the
equation of state by  Haensel, Zdunik and Dobaczewski (1989).
This equation of state (HZD)  is obtained in the following way:
The pressure is fitted
by a polynomial consisting of 9 terms, i.e.,
\begin{equation}
P(X)=\sum_{i=1}^9C_iX^{l_i},
\end{equation}
\noindent
where
\begin{equation}
X=1.6749\times10^5n,
\end{equation}
\par\noindent
$n$ is given in [fm$^{-3}$], and the values
\[
n= 0.077,~~ 0.154,~~ 0.395,~~ 0.762,~~ 1.575,
3.147,~~ 6.443,~~ 12.240,~~ 26.551,
\]
in [$10^{-5}$/fm$^3$], are chosen to give the coefficients $C_i$.
The corresponding equations are solved by matrix inversion, and we obtain
$$
P(X)=8.471521942X^{1/3}-40.437728191X^{2/3}+74.927783479X
$$
$$
-67.102601796X^{4/3}+30.011422630X^{5/3}-4.207322319X^{2}
$$
\begin{equation}
-1.419954871X^{7/3}+0.589441363X^{8/3}-0.060468689X^{3},
\end{equation}
where $P(X)$ is given in units of
[$10^{27}$ N/m$^2$] and in the density range of
$2\times10^{-6}$ fm$^{-3}<n< 2.84\times 10^{-4}$ fm$^{-3}$.
We need all the decimals in the different terms to get an accuracy of
at least two decimals in the net equation.

The  Baym-Bethe-Pethick (BBP)
equations of state (Baym et al. 1971) are taken from
{\O}verg{\aa}rd and {\O}stgaard (1991)
The given data are fitted by two
five-term polynomials to give (BBP-1)
$$
P(n)=4.3591n^{4/3}-122.4841n^{5/3}+1315.2746n^{2}
$$
\begin{equation}
-6180.0702n^{7/3}+10659.0049n^{8/3},   \label{eq:9}
\end{equation}
where $P(n)$ is given  in units of [$10^{34}$ N/m$^2$] for $n$
 in the density range of
$0.00027$ fm$^{-3} <n <0.0089$ fm$^{-3}$,
and ( BBP-2)
$$
P(n)=0.092718n^{4/3}-0.035382n^{5/3}+1.193525n^{2}
$$
\begin{equation}
-2.424555n^{7/3}+2.472867n^{8/3},
\end{equation}
where $P(n)$ is given  in units of [$10^{34}$ N/m$^2$] for $n$
in the density range of
$0.0089$ fm$^{-3} <n <0.3$ fm$^{-3}$.

The  Arntsen- {\O}verg{\aa}rd
(A{\O}-5)
equation of state is given by a five-term polynomial
(\O verg\aa rd \& \O stgaard 1991), i.e.,
$$
P(n)= 9.4433n^{5/3}-34.6909n^{2}+102.6575n^{8/3}
$$
\begin{equation}
-87.6158n^{3}+14.3549n^{11/3},
\end{equation}
where $P(n)$ is given  in units of [$10^{34}$ N/m$^2$] for $n$
in the density range of
$0.4$ fm$^{-3} <n <3.6$ fm$^{-3}$.

The  equation of state by
Pandharipande \& Smith (1975) (PS) is taken from
{\O}verg{\aa}rd and {\O}stgaard (1991). The given data are fitted by a
five-term polynomial to give
$$
P(n)=4.0378n^{4/3}-27.853n^{5/3}+52.0859n^{2}
$$
\begin{equation}
-20.7073n^{7/3}+5.5808n^{8/3},
\end{equation}
where $P(n)$ is given  in units of [$10^{34}$ N/m$^2$] for $n$
in the density range of
$0.1$ fm$^{-3} <n <3.6$ fm$^{-3}$.

For our non-relativistic  equations
of state   we find that the
following equations of state are the best to cover the whole
range of  densities in a neutron star, and  we use:

\noindent
HZD in the density range of
\[
 n< 0.000256,
\]
\noindent
BBP-1 in the density range of
\[
0.000256\le n < 0.003892,
\]
\noindent
BBP-2 in the density range of
\[
 0.003892\le n<n_1,
\]
our non-relativistic equations of state in the density range of
\[
n_1 \le n<n_2,
\]
\noindent
A{\O}-5 in the density range of
\[
 n_2\le n<3.46,
\]
\noindent
and PS in the density range of
\[
n  \ge 3.46,
\]
where $n$  is given in  units of [fm$^{-3}$].
The coupling points (densities) $n_1$ and $n_2$ are here given by
\[
n_1=0.10, \ n_2=0.80,\mbox{ for \ 0\%\ protons,}
\]
\[
n_1=0.20, \ n_2=0.85,\mbox{ for 5\%\ protons,}
\]
\[
n_1=0.20, \ n_2=0.87,\mbox{ for 10\%\ protons,}
\]
\[
n_1=0.35, \ n_2=0.90,\mbox{ for 15\%\ protons,}
\]
\[
n_1=0.40, \ n_2=1.05,\mbox{ for 20\%\ protons,}
\]
\[
n_1=0.50, \ n_2=1.15,\mbox{ for 25\%\ protons.}
\]
For our relativistic equation of state, we have coupled
the following equations of state:

\noindent
HZD in the density range of
\[
 n< 0.000256,
\]
\noindent
BBP-1 in the density range of
\[
0.000256 \le n< 0.003892,
\]
\noindent
BBP-2 in the density range of
\[
 0.003892 \le n<n_1,
\]
our relativistic equations of state in the density range of
\[
n_1\le n<n_2,
\]
\noindent
and PS in the density range of
\[
n  \ge n_2,
\]
where $n$  is given in units of [fm$^{-3}$].\newline
The coupling points (densities) $n_1$ and $n_2$ are here given by
\[
n_1=0.11, \ n_2=0.85,\mbox{ for \ 0\%\ protons,}
\]
\[
n_1=0.14, \ n_2=0.80,\mbox{ for 5\%\ protons,}
\]
\[
n_1=0.15, \ n_2=0.85,\mbox{ for 10\%\ protons,}
\]
\[
n_1=0.17, \ n_2=0.85,\mbox{ for 15\%\ protons,}
\]
\[
n_1=0.18, \ n_2=0.95,\mbox{ for 20\%\ protons,}
\]
\[
n_1=0.19, \ n_2=0.93,\mbox{ for 25\%\ protons,}
\]
\[
n_1=0.20, \ n_2=0.93,\mbox{ for 30\%\ protons,}
\]
\[
n_1=0.21, \ n_2=0.85,\mbox{ for 35\%\ protons,}
\]
\[
n_1=0.22, \ n_2=0.90,\mbox{ for 40\%\ protons,}
\]
\[
n_1=0.23, \ n_2=0.95,\mbox{ for 45\%\ protons,}
\]
\[
n_1=0.24, \ n_2=0.85,\mbox{ for 50\%\ protons.}
\]

These equations are chosen among 12 published equations of state,
and they seem to be  the best ones coupled together
in the total density range.
Total masses,  radii, moments of inertia and surface gravitational
redshifts are then calculated, and  parameterized as  functions of
 the central density $n_c$.
Fig.~2  shows  the total mass  and fig.~3 the radius versus the   central
density.
  Fig.~4 shows the mass versus the radius. Fig.~5 shows  the moment of inertia
and fig.~6 the gravitational redshift
versus the  total mass of the  star.

\section{Discussions and conclusions}

{}From figs.~2--4 we find from the non-relativistic EOS a  maximum mass
at  $M_{\mathrm{max}}= 1.65 M_{\odot}$ and the corresponding
radius $ R= 8.7 $ km  for pure neutron matter.
For a proton fraction of 25\% , we find the maximum mass to be
$M_{\mathrm{max}}= 1.37 M_{\odot}$ with a corresponding
radius $ R= 6.9 $ km.
The respective central densities are $1.65$ fm$^{-3}$ and
$2.33$ fm$^{-3}$ (i.e. an order of magnitude higher than normal
nuclear matter density).

The maximum mass calculated
relativistic equation of state for neutron matter yields  $M_{\mathrm{max}}=
2.38 M_{\odot}$
with  the corresponding radius  $R = 12.3$ km.
Increasing the proton fraction to  45 \% ,
we get a maximum mass of $M_{\mathrm{max}}= 2.07 M_{\odot}$
with the corresponding radius  $R = 10.3$ km.
The respective central densities are $0.75$ fm$^{-3}$ and
$0.95$ fm$^{-3}$.

We see that  stars calculated with stiff
equations of state have greater maximum mass, lower central density
(and thicker crust) than stars obtained with soft equations of state.

Observations of binary pulsars give  maximum neutron star masses of
(\O verg\aa rd \& \O stgaard 1991; Prakash et al. 1988)
\[
1.0M_{\odot} < M_{\mathrm{max}} < 2.2 M_{\odot},
\]
\noindent or possibly
(Thorsett et al. 1993; Finn 1994; Taylor \& Weisberg 1989;
Joss \& Rappaport 1984; Glendenning 1988)
\[
1.3M_{\odot} < M_{\mathrm{max}} < 1.85 M_{\odot},
\]
\noindent
At present, no reliable measurements of the  radius of a neutron star exist.
But general estimates give
(\O verg\aa rd \& \O stgaard 1991)
\[
R \approx 9 \mathrm{km}.
\]
\noindent If this estimate is close to the true value, then the results
from our non-relativistic equations of state may look more
reasonable than those from the  relativistic one. However,
theoretical calculations of the radius of neutron stars
can not be confirmed very well by observational data, and  are more dependent
than the total mass on the  low-density equation of state. Moreover the maximum
mass
occurs at a very high central density, where  the relativistic effects
certainly
prevail.

Data on the nuclear equation of state can, in principle, be obtained
from several different sources such as the monopole resonance in nuclei,
 high energy nuclear collisions, supernovae and neutron stars.
Until recently it has, for instance, been assumed that the compression modulus
was reasonably well known from the analysis of the giant monopole resonance
in nuclei
(Blaizot et al. 1976; Blaizot 1980; Treiner et al. 1981).
 Later, however, these results were
questioned by Glendenning (1988) and Brown \& Osnes (1985).

Supernova simulations seem to require an equation of state which is too
soft to support some observed masses of neutron stars, if sufficient
 energy shall be released to make
the ejection mechanism work
(Baron et al. 1985; Woosley \& Weaver 1986; Baron et al. 1987).

Supernova explosions can then probably not give a reliable
constraint on the nuclear equation of state. Some analyses of high
 energy nuclear collisions, however, have indicated a moderately stiff
or very stiff equation of state
(Sano et al. 1985; Stocker \& Greiner 1986; Molitoris et al.1985),
although ambiguities have been observed by
Gale et al. (1987) and Sharma et al. (1987).
Various nuclear data and neutron star masses then seem to favour a rather high
 compression modulus of $K\sim 300$ MeV
(Glendenning 1988; Sharma et al. 1988; Sharma et al. (1989).
 No definite statements can be made, however, about the equation of state at
high densities, except that  the neutron star equation of
state  should probably be moderately stiff to support neutron star masses
 up to approximately 1.85 $M_\odot$ (Thorsett et al. 1993).

With the above observations, it seems that our
relativistic EOS for neutron matter is too stiff, since the predicted mass
$M_{max}\approx 2.37M_\odot$ and radius are larger than the estimated values.
We have found that nuclear matter including electrons and muons in beta
equilibrium
results in a softer EOS. The proton fraction in beta equilibrium is
approximately 10-15\%.

However, several mechanisms can  soften the equation of state for
neutron stars
considerably.
Among these we have mentioned in the introduction  kaon  and  pion
condensation.
Condensation of the negative charged mesons may be  likely to occur if the
chemical
potential of the mesons becomes  equal to the electron chemical potential.
If this situation occurs, the proton abundance would increase considerably,
possibly up to more than $40\%$ protons, and produce a softer EOS.
As mentioned above, when increasing  the proton fraction to 45\% , the
calculated maximum mass is reduced to approximately $ 2M_{\odot}$
with a corresponding radius of $R\approx 10$ km.
This  is still slightly higher than the experimental values
(Thorsett et al. 93).
It should, however, be noted that the gain in condensation energy
obtained due to the phase transition to a meson condensate will
bring the value of the maximum mass further down.

Further  processes
 which can soften the  equation of state
are conversion of nucleons to  hyperons or a phase transition
to quark matter at high densities, which  would  lower the energy due to an
increase of  the number of degrees of freedom
(Drago et al. 1995).
However, for a neutron star to resist  the centrifugal forces from very
fast rotation, the equation of state should be soft at low and
intermediate densities  and stiff at high densities, which would not
fit very well with the concept of  quark matter in
 hybrid stars (Weber et al. 1991).

{}From Figs.\ 2 and 4 we see that the relativistic equations of state
  give  minimum masses
$M_{\mathrm{min}}\approx 1.65  M_{\odot}$ at central densities around 0.4-0.5
fm$^{-3}$.
This result is almost independent of the proton fraction.
This is somewhat  higher than the  measured masses of neutron stars from
ref. (Thorsett et al. 1993).
{}From this we can conclude  that the  relativistic EOS needs additional
softening
to fit the data discussed.
Even if we assume our  relativistic EOS to be a good
approximation  at normal nuclear matter densities, there have to be softening
effects
present  at $n_c<0.5$ fm$^{-3}$.

In connection to this we stress that our calculation
of the EOS is to first order in the reaction matrix $G$, and we would
expect higher-order many-body contributions to also soften the EOS.
This was indeed shown in a preliminary study for
symmetric nuclear matter by Jiang et al. (1993).
Although only a set of higher-order contributions was considered, these
authors obtained a softening of the relativistic EOS. Such
effects will be studied by us in future works.

{}From fig.~5 we see that
our non-relativistic equations of state  give  values
for the moment of inertia of
\[
I(M_{\mathrm{max}})= 1.05 \times 10^{45} \mathrm{g cm^2},
\]
\[
I(M_{\mathrm{max}})=0.59 \times 10^{45} \mathrm{g cm^2},
\]
\noindent
at the  maximum mass for pure neutron matter and 25\% protons, respectively.
\noindent
The maximum values for moments of inertia obtained from
non-relativistic equations of state are
\noindent
\[
I_{\mathrm{max}}=1.16 \times 10^{45} \mathrm{g cm^2},
\]
\noindent
\[
I_{\mathrm{max}}=0.65 \times 10^{45} \mathrm{g cm^2},
\]
\noindent
for pure neutron matter and 25\% protons, respectively.
The relativistic EOS give
\noindent
\[
I(M_{\mathrm{max}})= 3.17 \times 10^{45}\mathrm{ g cm^2},
\]
\noindent
\[
I(M_{\mathrm{max}})= 2.29 \times 10^{45}\mathrm{ g cm^2},
\]
\noindent
for pure neutron matter and 45\% protons, respectively, and
maximum values for moments  of inertia are
\noindent
\[
I_{\mathrm{max}}=3.47 \times 10^{45} \mathrm{g cm^2},
\]
\noindent
\[
I_{\mathrm{max}}=2.39 \times 10^{45} \mathrm{g cm^2}.
\]
\noindent
These values  are not contradictory to  observations, and are
consistent with
the expansion of the Crab nebula and the luminosity  and the loss of
rotational energy from the Crab pulsar ( M\o lnvik 1985).

{}From fig.~6 we see that the gravitational surface redshift is not
strongly affected by the different equations of state. This is because
the density profiles of the stars are such that their
surface gravities are almost the same. A measurement of the redshift
can therefore not be used to distinguish between different types of stars or
equations of state. It is, however, possible that the slowing down of pulsars
and the corresponding glitches can give some
information about the internal structure.

In summary, in this work we have calculated the EOS for asymmetric nuclear
matter using different proton fractions.
Both a non-relativistic and a relativistic Brueckner-Hartree-Fock
procedure were employed in order to derive the equation of state,
which is the basic input quantity
in  neutron star calculations since it connects
the nuclear physics and the astrophysics. Of importance here is the fact
that a relativistic nuclear matter calculation with the Bonn A potential
meets the empirical nuclear matter data, a feature not accounted
for by non-relativistic calculations.
 By varying the
proton fractions we have estimated certain  limits of the neutron star
observables from both
a relativistic and a non-relativistic approach.
{}From the relativistic approach we obtain a maximum mass of the neutron star
that is slightly higher
than the empirical ones even when we have a proton fraction close to symmetric
nuclear  matter.
The relativistic effects become important at densities around and higher
than the saturation density for nuclear matter, and their main effect is
to stiffen the EOS at these densities. This mechanism is due to the fact that
the relativistic effective mass of the nucleon becomes smaller compared
to the free mass, an effect which in turn enhances the repulsive spin-orbit
term.
We obtain a rather low central density which implies that our EOS is too stiff.
However, we have discussed
several effects that can bring the relativistic  results closer
to the empirical values.
{}From the non-relativistic EOS  we obtain masses and radii  closer to
empirical values.
However, this approach yields an EOS  that is too soft to reproduce
the empirical data of symmetric nuclear matter.
Furthermore, the non-relativistic approach gives substantial
proton superconductivity in the interior of the neutron star
(De Blasio et al. 1995),
and thereby inhibiting the traditional URCA mechanisms.

{}From our investigation we conclude that  our relativistic equations
of state are too stiff to reproduce neutron star properties like
mass and radius.
The existence of  exotic states of nuclear matter,
such as kaon or pion condensate or quark matter,  may explain
these discrepancies.
However, we may also expect the relativistic EOS to be softened by
higher-order many-body contributions.

The other observables like moments of inertia and
gravitational redshifts are in good agreement with the accepted values for
both the non-relativistic and the relativistic approach.

This work has been supported by the NorFa (Nordic Academy for Advanced
Study).
MHJ thanks the Istituto Trentino di Cultura, Italy, and the Research
Council of Norway (NFR) for financial support. The calculations have been
carried out at the IBM cluster at the University of Oslo. Support
for this from the NFR is acknowledged.

\begin{figure}[h]
\caption{Energy per particle for asymmetric nuclear matter
as function of particle density for
different proton fractions. Relativistic results are indicated by solid lines
and non-relativistic by dashed lines. The increments in proton fractions are
5\% for both kinds of results going
from pure neutron matter for the top solid and dashed  curves
to symmetric nuclear matter for the corresponding lowest curves.
}
\label{fig:energy}
\end{figure}

\begin{figure}[h]
\caption{
Total mass in units of solar masses [$M_\odot$] as function of  central density
for neutron stars.
Results obtained from  relativistic equations of state are indicated by solid
lines,
and results obtained from non-relativistic EOS are indicated by dashed lines.
The results are  shown
for different proton fractions.
}
\label{fig:fig2}
\end{figure}
\begin{figure}[h]
\caption{
Total radius as function of  central density for neutron stars.
Results obtained from relativistic EOS are  indicated by solid lines,
and  results obtained from  non-relativistic EOS are indicated by dashed lines.
The results are shown
for different proton fractions.
}
\label{fig:3}
\end{figure}
\begin{figure}
\caption{
Mass-radius relations $M(R)$ for neutron stars.
For further explanations; see Fig. 1 and Fig. 2.
}
\label{fig:fig4}
\end{figure}
\begin{figure}
\caption{
Moment of inertia as function of  total mass given in [$M_\odot$] for neutron
stars.
Results obtained from relativistic EOS are indicated by solid lines,
and the results obtained from non-relativistic EOS are indicated by dashed
lines.
}
\label{fig:fig5}
\end{figure}
\begin{figure}
\caption{
Surface gravitational redshift as function of  total mass [$M_\odot$]
for neutron stars. For further explanations; see Fig. 5.
}
\label{fig:fig6}
\end{figure}
\clearpage
Aichelin J.,  Rosenhauer A., Peilert G., Stocker H. \&
Greiner W., Phys.\ Rev.\ Lett.\ {\bf 58} (1987) 1926  \newline
Bao G., Engvik L., Hjorth-Jensen M.,  Osnes E.
\& \O stgaard E., Nucl.\ Phys.\ {\bf A 575} (1994) 707  \newline
Baron E., Bethe H. A.,  Brown G. E.,  Cooperstein J.
 \& Kahana S., Phys.\ Rev.\ Lett.\ {\bf 59} (1987) 736    \newline
Baron E.,  Cooperstein J. \&
Kahana S., Phys.\ Rev.\ Lett.\ {\bf 55}
(1985) 126    \newline
Baym G.,  Bethe H. A. \& Pethick C. J.,
Nucl.\ Phys.\ {\bf A175} (1971) 225    \newline
Blaizot J. P., Phys.\ Rep.\ {\bf 64} (1980)171  \newline
Blaizot J. P.,  Gogny D.  M.K. \&  Grammaticos B., Nucl.\ Phys.\
{\bf A265} (1976) 315      \newline
Brockmann R., Phys.\ Rev.\ {\bf C18} (1978) 1510    \newline
Brockmann R. \& Machleidt R.,
Phys.\ Rev.\ {\bf C42} (1990) 1965    \newline
Brown G. E., in proceedings of Perspectives in Nuclear
Structure, Niels Bohr Institute, June 14-19, 1993, Nucl.\ Phys.\
{\bf A574} (1994) 217c    \newline
Brown G.E., Lee C.-H., Rho M. \& Thorsson V.,
Nucl.\ Phys.\ {\bf A567} (1994) 937 \newline
Brown G. E. \& Osnes E., Phys.\ Lett.\ {\bf B159} (1985) 223    \newline
Brown G. E.,  Weise W.,  Baym G. \&  Speth J.,
Comments on Nucl.\ and Part.\ Physics {\bf 17} (1987) 39    \newline
Casares J.,  Charles P. A. \&  Naylor T.,
Nature {\bf 355} (1992) 614    \newline
Celenza L. S. \& Shakin C. M., {\em Relativistic
Nuclear Physics: Theories of Structure and Scattering}, Vol. 2
of Lecture Notes in Physics (World Scientific, Singapore, 1986)     \newline
Cowley A. P.,  Crampton D.,  Hutchings J. B.,
Remillard R. A. \& Penfold J. P., Astrophys.\ Journ.\ {\bf 272} (1983) 118
\newline
De Blasio F. V.,  Elgar\o y \O .,  Engvik L.,
Hjorth-Jensen M., Lazzari G. \&  Osnes E. in preparation    \newline
Dickhoff W. H. \& M\"{u}ther H., Rep.\ Prog.\ Phys.\
{\bf 55} (1992) 1947    \newline
Drago A.,   Tambini U. \& Hjorth-Jensen M.,
ECT* preprint ECT/MAY/95-03 and submitted to Phys.\ Rev.\ Lett.    \newline
Engvik L.,  Hjorth-Jensen M.,   Osnes E.,
Bao G. \& \O stgaard E.,
Phys.\ Rev.\ Lett.\ {\bf 73}, (1994) 2650    \newline
Finn L. S.,  Phys.\ Rev.\ Lett. {\bf 73} (1994) 1878  \newline
Gale C., Bertsch G. \&  Das Gupta S., Phys.\ Rev.\ {\bf C35}
(1987) 415    \newline
Gies D. R. \&  Bolton C. T.,
Astrophys.\ Journ.\ {\bf 304} (1986) 371    \newline
Glendenning N. K., Phys.\ Rev.\ {\bf C37} (1988) 2733    \newline
Glendenning N. K.
in Proceedings of the International Summer School on
``Structure of Hadrons and Hadronic
Matter'' (World Scientific, Singapore, 1991), 275    \newline
Haensel P.,  Zdunik J. L. \& Dobaczewski J.,
Astron.\ Astrophys.\ {\bf 222} (1989) 353    \newline
Horowitz C. J. \& Serot B. D., Nucl.\ Phys.\ {\bf A464}
(1987) 613    \newline
Itzykson C. \&  Zuber J.-B. {\em Quantum Field theory}
(McGraw-Hill, New York, 1980) \newline
Jiang M. F.,  Machleidt R.  \&  Kuo T. T. S.,
Phys.\ Rev.\ C {\bf 47} (1993)  2661   \newline
Joss P. C. \&  Rappaport S. A., Ann.\ Rev.\ Astron.\ Astrophys.\
{\bf 22} (1984) 537    \newline
Kuo T. T. S. \& Ma Z. Y.,
Phys.\ Lett.\ {\bf B127} (1983) 123 \newline
Kuo T. T. S. \&  Ma Z. Y., in
{\em Nucleon-Nucleon Interaction and Nuclear
Many-body Problems}  (edited by  Wu S. S. and  Kuo T. T. S.)
(World Scientific, Singapore) p. 178     \newline
Kuo T. T. S.,  Ma Z. Y.  \&  Vinh Mau R.,
Phys.\ Rev.\ {\bf C33} (1986) 717    \newline
Lattimer J. M.,  Pethick C. J., Prakash M. \&
Haensel P., Phys.\ Rev.\ Lett.\ {\bf 66} (1991) 2701 \newline
Li G.Q.,  Machleidt R. \&  Brockmann R.,
Phys.\ Rev.\ {\bf C46} (1992) 2782    \newline
Lorenz C. P.,  Ravenhall D. G. \& Pethick C. J.,
Phys.\ Rev.\ Lett.\ {\bf 70} (1993) 379     \newline
Machleidt R., Adv.\ Nucl.\ Phys.\ {\bf 19} (1989) 189 \newline
Machleidt R. \&  Li G. Q., Phys.\ Rep.\ {\bf 242} (1994) 5    \newline
Mc Clintock J.E. \& Remillard R. A.,  \newline
Astrophys.\ Journ.\ {\bf 308} (1986) 110
Mahaux C., Bortignon P. E.,  Broglia R.A. \&  Dasso C. H.,
 Phys.\ Rep.\ {\bf 120} (1985) 1    \newline
Mahaux C. \& Sartor R.,
Phys.\ Rev.\ {\bf C 40} (1989) 1833    \newline
Migdal A. B., Saperstein E. E., Troitsky M. A.
\& Voskresensky D. N., Phys.\ Rep.\ {\bf 192} (1990) 179 \newline
Mittet R. \&  \O stgaard E., Phys.\ Rev.\ C {\bf 37} (1988) 1711  \newline
Molitoris J. J.,  Hahn D. \& Stocker H., Nucl.\ Phys.\ {\bf A447}
(1985) 13c    \newline
Muto T.,  Takatsuka T., Tamagaki R.
\& Tatsumi T., Prog.\ Theor.\ Phys.\ Suppl.\ {\bf 112} (1993) 221    \newline
Muto T., Tamagaki R.
\&  Tatsumi T., Prog.\ Theor.\ Phys.\ Suppl.\ {\bf 112} (1993) 159    \newline
M\"{u}ther H.,  Machleidt R. \& Brockmann R., Phys.\ Rev.\
{\bf C42} (1990) 1981    \newline
M{\o}lnvik T. \& {\O}stgaard E.,  Nucl.\ Phys.\ {\bf A437}
(1985) 239 \newline
Nikolaus T.,  Hoch T.  \& Madland D. G., Phys.\ Rev.\
{\bf C46} (1992) 1757 \newline
Pandharipande V. R.  \& Smith R. A.
Nucl.\ Phys.\ {\bf A237} (1975) 507 \newline
Pethick C.\ J.\ \& Ravenhall D.\ G., Phil.\ Trans.\
Roy.\ Soc.\ Lond.\ {\bf A341} (1992) 17    \newline
Pethick C. J.,  Ravenhall D. G. \& C.\ P.\ Lorenz C. P.,
Nucl.\ Phys.\ {\bf A584} (1995) 675    \newline
Prakash M.,  Ainsworth T. L. \&
Lattimer J. M., Phys.\ Rev.\ Lett.\
{\bf 61} (1988) 2518    \newline
Ray L., Hoffman G.W. \& Cooker W. R., Phys. Rep. {\bf 212}
(1991) 223    \newline
Rosenhauer A., Staubo E. F.,  Csernai L. P.,
 \O verg\aa rd T. \&  \O stgaard E., Nucl.\ Phys.\ {\bf A540} (1992) 630
\newline
Sano M., Gyulassy M., Wakai M. \& Kitazoe Y.,
Phys.\ Lett.\ {\bf B156} (1985) 27    \newline
Sharma M. M., Borghols W. T. A., Brandenburg S., Crona S.,
van der Woude A. \&  Harakeh M. N., Phys.\ Rev.\ {\bf C38} (1988) 2562
\newline
Sharma M. M.,  Stocker W.,  Gleissl P. \& Brack M.,
Nucl.\ Phys.\ {\bf A504} (1989) 337    \newline
Serot B. D., Rep.\ Prog.\ Phys.\ {\bf 55} (1992) 1855    \newline
Serot S. D. \& Walecka J. D.,
Adv.\ Nucl.\ Phys.\ {\bf 16}  (1986) 1    \newline
Stocker H. \&  Greiner W., Phys.\ Rep.\ {\bf 137} (1986) 277 \newline
Takatsuka T., Tamagaki R.
\& Tatsumi T., Prog.\ Theor.\ Phys.\ Suppl.\ {\bf 112} (1993) 67    \newline
Taylor J. H. \&  Weisberg J. M.,
Astrophys.\ Journ.\ {\bf 345} (1989) 434    \newline
ter Haar B.  \& Malfliet R., Phys. Rep. {\bf 149} (1987) 207 \newline
Thorsett S. E.,  Arzoumanian Z.,  McKinnon M. M.
\& Taylor J. H., Astrophys.\ J.\ {\bf 405} (1993) L29    \newline
Thompson R. H., Phys.\ Rev.\ {\bf D1} (1970) 110    \newline
Thorsson V.,  Prakash M. \&  Lattimer J. M.,
Nucl.\ Phys.\ {\bf A572} (1994) 693    \newline
Treiner J.,  Krevine H.,  Bohigas O.  \& Martorell J.,
Nucl. Phys. {\bf A317} (1981) 253    \newline
Weber F.\ \&  Glendenning N.\ K., ``Hadronic Matter
and Rotating Relativistic Neutron Stars'', in Proceeedings of the
Nankai Summer School on Astrophysics and Neutrino physics
(World Scientific, Singapore, 1993),  64-183. \newline
Weber F.,  Glendenning N. K. \&  Weigel M. K.,
Astrophys. Journ. {\bf 373} (1991) 579    \newline
Wiringa R.B., Fiks V. \& Fabrocini A., Phys.\ Rev.\
{\bf C38} (1988) 1010    \newline
Witten E., Phys.\ Rev.\ {\bf D 30} (1984) 272    \newline
Woosley S.E. \& Weaver T. A., Ann.\ Rev.\ Astron.\ Astropys.\
{\bf 24} (1986) 205    \newline
{\O}verg{\aa}rd T. \&  {\O}stgaard E.,
Can.\ Journ.\ Phys.\ {\bf 69} (1991) 8 \newline

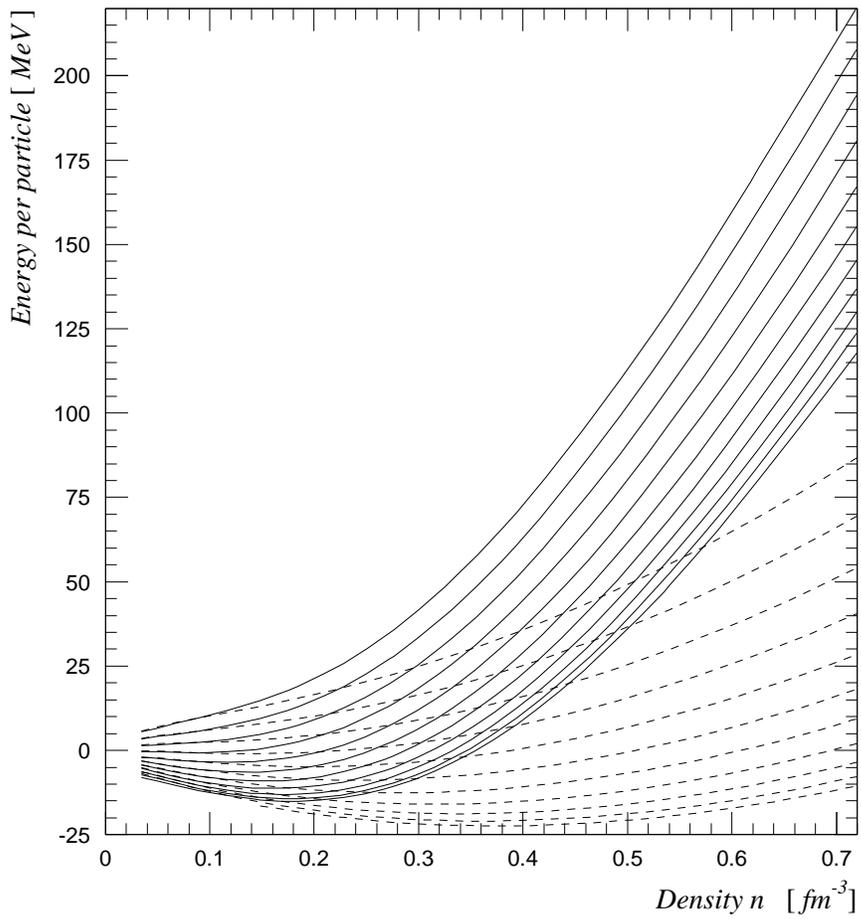
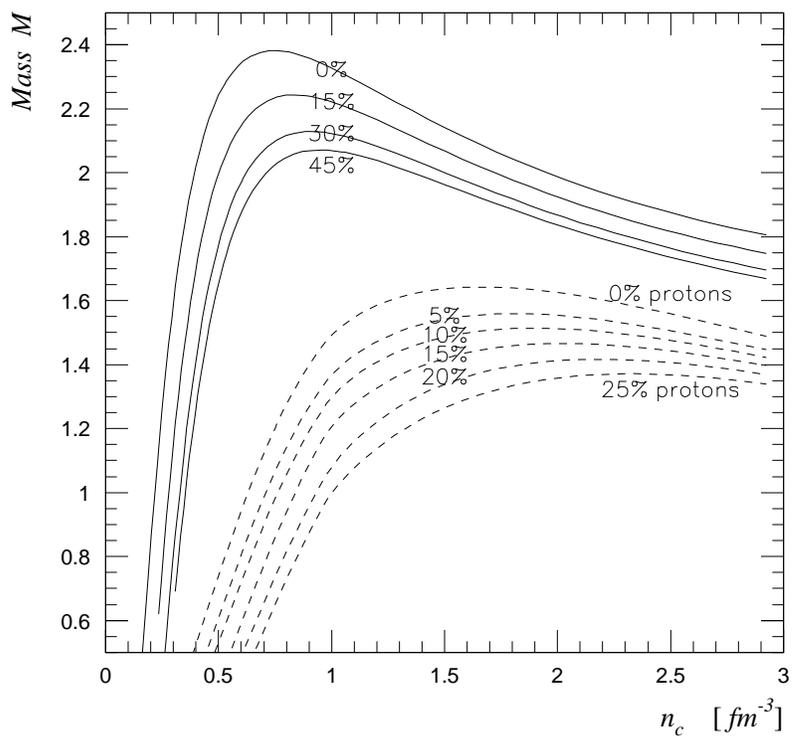
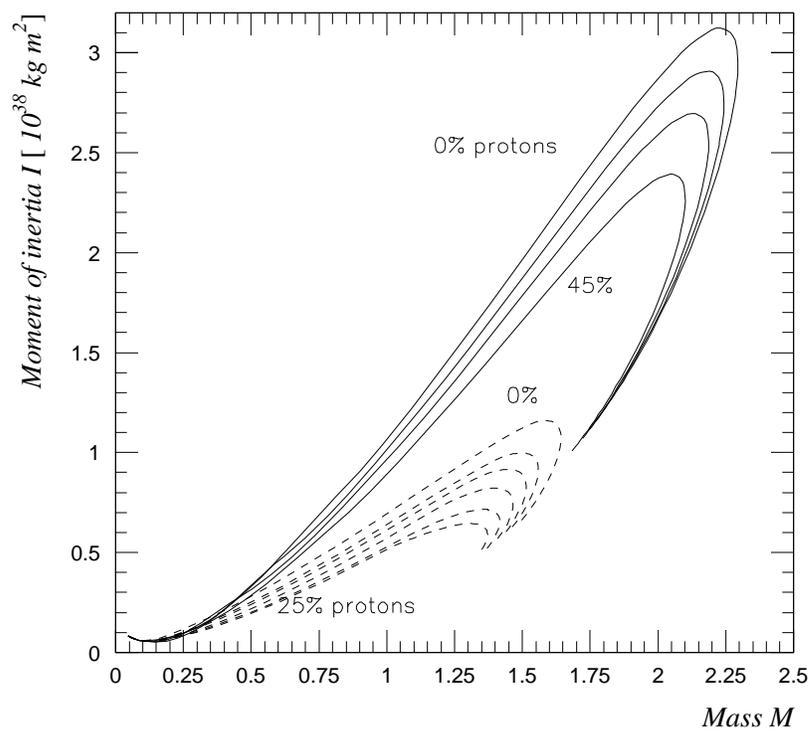
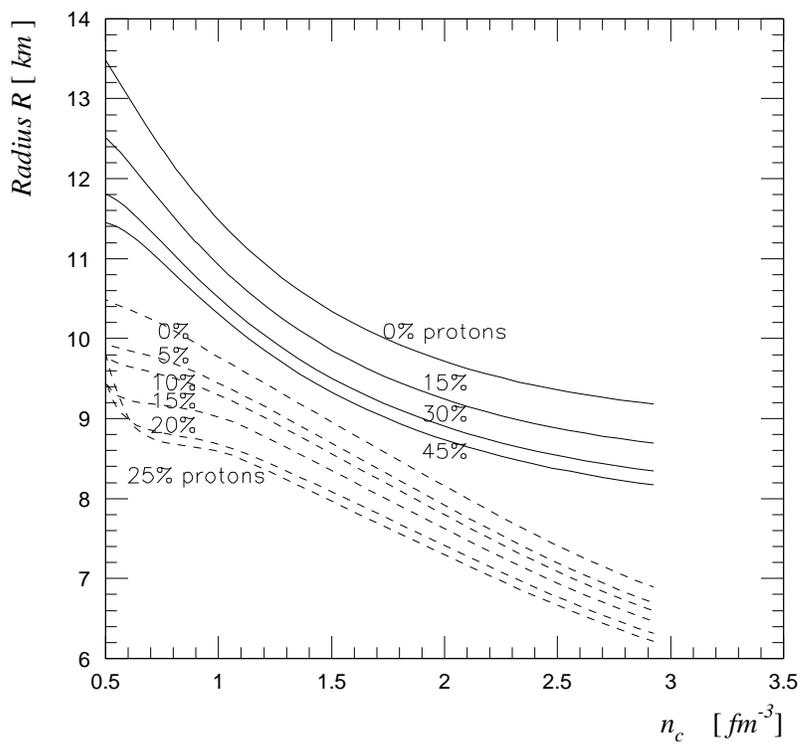
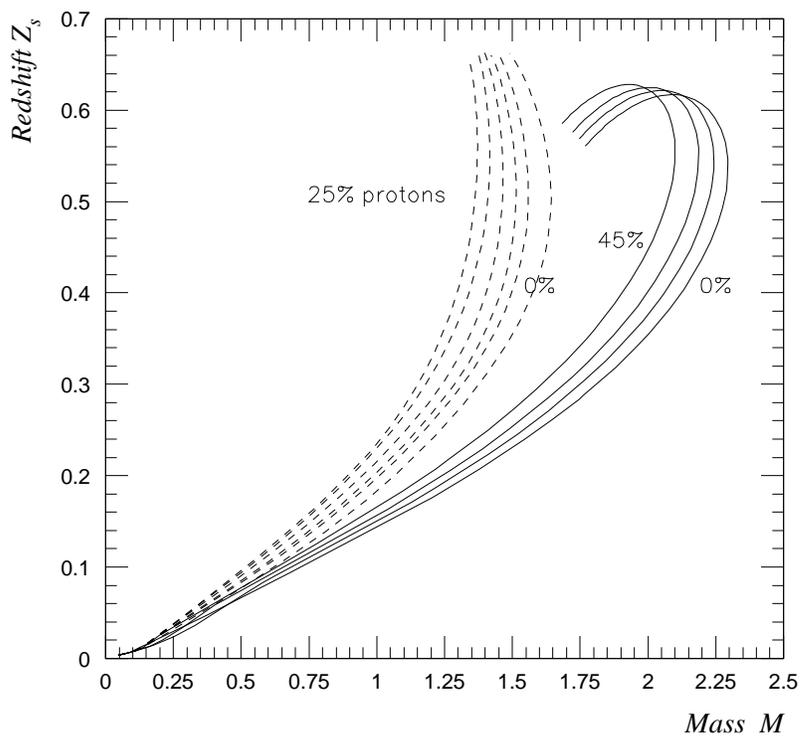
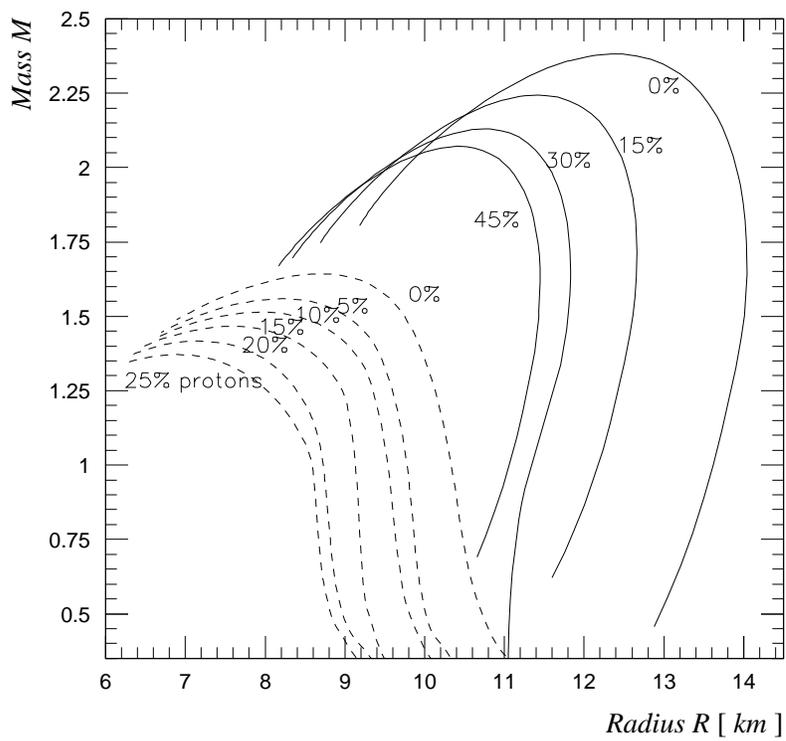
\end{document}